\documentclass{article}
\usepackage{ijcai25}

\usepackage{times}
\usepackage{soul}
\usepackage{url}
\usepackage[hidelinks]{hyperref}
\usepackage[utf8]{inputenc}
\usepackage[small]{caption}
\usepackage{graphicx}
\usepackage{amsmath}
\usepackage{amsthm}
\usepackage{booktabs}

\usepackage[switch]{lineno}
\usepackage{amsthm,amsmath,amssymb}
\usepackage{mathrsfs}

\usepackage{algorithm}
\usepackage{algpseudocode}

\usepackage{caption}
\usepackage{graphicx}
\usepackage{float} 
\usepackage{subcaption}
\usepackage{tcolorbox}
\usepackage{diagbox}
\usepackage{multirow}
\usepackage{lipsum}
\usepackage{makecell}
\usepackage{pifont}
\usepackage{engord}
\usepackage{microtype}
\usepackage{flushend}
\usepackage{balance}
\usepackage{appendix}

\title{Robust Multi-agent Communication Based on Decentralization-Oriented Adversarial Training}
\author{
Xuyan Ma$^{1,2,3}$
\and
Yawen Wang$^{1,2,3}$ \thanks{Corresponding authors}
\and
Junjie Wang$^{1,2,3, *}$
\and
Xiaofei Xie$^4 $\and
Boyu Wu$^{1,2,3}$\And
Shoubin Li $^{1,2,3}$\And
Fanjiang Xu $^{1,2,3}$\And
Qing Wang $^{1,2,3, *}$
\\
\affiliations
$^1$State Key Laboratory of Intelligent Game, Beijing, China 
\\
$^2$Science and Technology on Integrated Information System Laboratory, Institute of Software Chinese Academy of Sciences, Beijing, China
\\
$^3$University of Chinese Academy of Sciences, Beijing, China\\
$^4$Singapore Management University\\
\emails
\{maxuyan2021, yawen2018, junjie, shoubin, fangjiang, wq\}@iscas.ac.cn,
xfxie@smu.edu.sg,
boyu\_wu2021@163.com \\
}

\usepackage{soul}
\usepackage{enumitem}

\newcommand{\tool}{DMAC}
\newcommand{\tooladv}{DMAC\_Adv}

\newcommand{\website}{https://anonymous.4open.science/r/IJCAI2025-FB61}

\begin{document}
\maketitle

\begin{abstract}
In typical multi-agent reinforcement learning (MARL) problems, communication is important for agents to share information and make the right decisions. 
% At the same time, it has been shown that communication policies are vulnerable to adversarial attack, such as attacking communication channels and message. However, research on how to develop robust communication policy has been largely ignored. 
However, due to the complexity of training multi-agent communication, existing methods often fall into the dilemma of local optimization, which leads to the concentration of communication in a limited number of channels and presents an unbalanced structure. Such unbalanced communication policy are vulnerable to abnormal conditions, where the damage of critical communication channels can trigger the crash of the entire system.
% At the same time, adversarial machine learning (ML) has shown that ML models are vulnerable to attack.
% \yawen{comment}
% 为啥是ML models vulnerable？这篇文章关注的是啥？这句话跟前后逻辑不顺，不如直接说通讯模型容易vulnerable，例如xx干扰，但是对于如何提升robust的相关研究很少
% 现在的写法还是在按照提升robust的思路写的，但是按照题目的说法，我们是要按照decentral的思想提出一个通讯模型，那这样是不是跟robust没关系了？所以这些robust的说法还要保留吗?而且后面rq1 是不是就应该是性能了，然后再是robust？这个需要再确认下
% Despite increasing attention to the robustness of MARL algorithms, how to achieve robust communication policy %in multi-agent reinforcement learning 
% has been largely ignored. 
Inspired by decentralization theory in sociology, we propose {\tool}, which enhances the robustness of multi-agent communication policies by retraining them into decentralized patterns.
Specifically, we train
% \rev{construct} 
an adversary {\tooladv} which can dynamically identify and mask the critical communication channels, and then apply the adversarial samples generated by {\tooladv} to the adversarial learning of the communication policy to force the policy in exploring
% \rev{to explore} 
other potential communication schemes and transition to a decentralized structure.
% \xuyan{comment} %reward相关的是不是可以不放到摘要里
As a training method to improve robustness, {\tool} can be fused with any learnable %-based 
communication policy algorithm.
% We model it as a MARL problem to capture the relationships and communications between agents. 
% \yawen{comment}
% 上面这两句话adversary和the cutting agents不是一个东西吗?合并一下吧，后面那个说明reward怎么做的这块，是不是想说明adversary怎么训练的？那跟到第一句后面吧，目前abstract正好也有些长
% \yawen{comment}
% cutting agent名字不太好，后面再想想。
% 这里还是没有说明白为啥构建了adversary {\tooladv}，然后再对抗训练，就可以实现decentralized
% Specifically, we define the input of the agents and reconstruct the reward function to minimize the gap between expected reward and actual reward. 
The experimental results in two communication policies and four multi-agent tasks demonstrate that {\tool} achieves higher improvement on robustness and performance of communication policy compared with two state-of-the-art and commonly-used baselines. Also, the results demonstrate that {\tool} can achieve decentralized communication structure with acceptable communication cost.
% the component {\tooladv} can correctly identify critical communication connections.
\end{abstract}
\section{Introduction}
\label{sec:intro}

Multi-agent reinforcement learning (MARL) has attracted more and more attention due to its powerful ability to solve complex problems, such as autonomous driving \cite{Shalev,SallabAPY17}, multi-agent control \cite{CenaCPPS13,DuanCX12} and unmanned aerial vehicles \cite{Zhiliang,WangWPXAH21}. 
In a multi-agent system (MAS), especially in a cooperative task, communication usually plays {a crucial role in facilitating coordination and enhancing overall performance. Effective communication enables agents to share information, make informed decisions, and synchronize their actions, ultimately leading to improved outcomes while weak communication may prevent agents from making safe and reliable decisions in the face of interference  \cite{ShengW0LWYCZ23}.}
% \rev{an important role. }
% \rev{MAS is typically trained in ideal communication conditions, but weak communication policies in real applications may prevent agents from making safe and reliable decisions due to interference.}

\begin{figure}[tbp]
    \centering
    \includegraphics[width=0.48\textwidth]{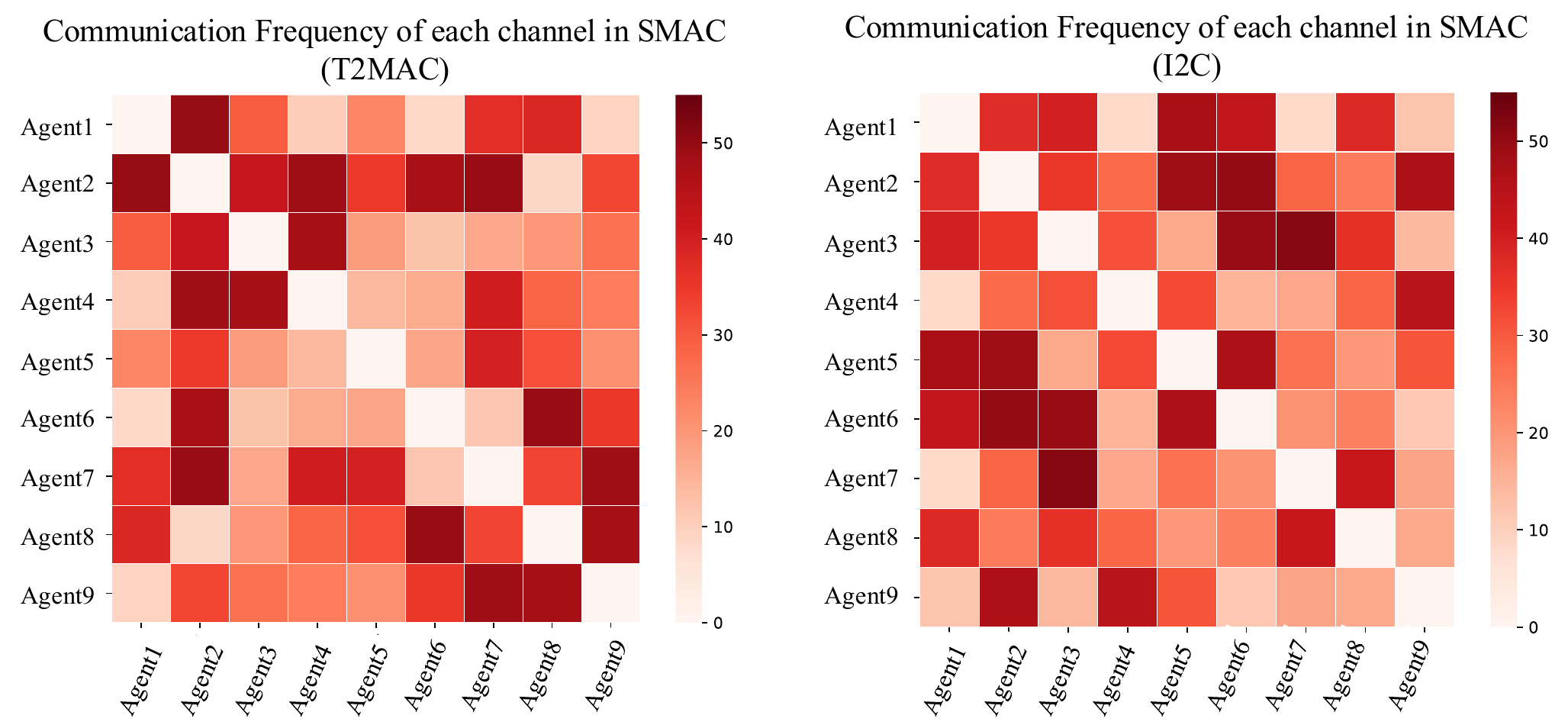}
    \caption{The communication frequency of the communication channel between each pair of agents in two environments. The darker the color, the more times the two agents communicate, that is, the higher the communication frequency of the communication channel. The grids on the diagonal represent the same agent and are therefore white.}
    % For intuition, the results are arranged in descending order. Due to the space limit, we omit the horizontal coordinate which represents all communication channels.}
    % An example of test-time communication attacks in a communicative MARL system. (a) During training, agents are trained collaboratively in a safe environment, such as a simulated environment. (b) In deployment, agents execute pre-trained policies in the real world, where malicious attackers may attack and cut off the connections to mislead some victim agent(s).
    \label{fig:intro}
% \vspace{-0.1in}
\end{figure}

In the early stage, communication is often carried out by broadcasting \cite{SukhbaatarSF16,DasGRBPRP19}, but such communication often produces redundant messages, which leads to a large amount of resource consumption. 
Later, researchers begin to focus on learning appropriate communication policies rather than broadcasting, which could improve communication efficiency and reduce communication redundancy by deciding whether or what to communicate between two agents \cite{ZhuD024}.
T2MAC\cite{SunZLLXWZ24} determines whether agents communicate with each other by quantifying the strength and relevance of each communication link between agents. I2C \cite{DingHL20} learns to decide whether to communicate with each of the other agents, by evaluating the effect of other agents on an agent’s own strategy.  
% \xuyan{MAS is usually trained under good communication conditions, and each agent gradually learns to trust all obtained communication messages and make use of them. However, when encountering interference in the real application, weak communication polices may cause agents to be unable to make safe and reliable decisions.} 
{However, developing an optimal communication policy is inherently challenging: existing approaches struggle to achieve comprehensive optimization and frequently become trapped in local optima, resulting in excessive communication concentrated among a limited number of agents and leading to an imbalanced communication structure.}
We count the communication frequency and find that the communication decisions generated by the existing methods did present an unbalanced distribution, as shown in Figure \ref{fig:intro} for T2MAC and I2C. The communication frequency of several communication channels is very high (darker color) and that of other channels is very low (lighter color). 
Take T2MAC as an example, 30\% of communication channels account for nearly 70\% of communication frequencies.
% \jie{comment}
%这里能加一句，大概百分之多少的智能体占据了百分之多少的通讯量或者之类的，例如 20% 占据了80%，让大家看出来这个事。现在这两个图吧，就是觉得有深的，有浅的，能看出来肯定是unbalanced，但看不出来多么的unbalance。 
{According to decentralization theory in sociology \cite{boss2023robustness}, a network that is overly centralized risks collapse if a few key nodes fail, while a more decentralized network exhibits greater resilience under attack. 
This highlights the vulnerability of unbalanced communication policies in MAS, where the failure of critical communication channels can lead to catastrophic breakdowns of the entire system.
}
% \rev{According to the decentralization theory in sociology (If a network is too centralized, the failure of a few key nodes may cause the system to collapse. In a more decentralized network, the system will show greater robustness if an attack occurs.),
% such an unbalanced communication policy is vulnerable to adversarial attack, such as the entire MAS will be crashed when important communication channels are attacked . }
% \xuyan{comment} %上面括号里的关于decentralization theory的介绍待定
% Subsequently, we counted the degree of each vertex (i.e., the number of connections between a vertex and its neighbors), as shown in the Figure \ref{fig:intro}. We can see that the number of communication connections presents an unbalanced and uneven distribution. This unbalanced communication architecture is often vulnerable to adversarial attack, so we want to change the current architecture to enhance the robustness of the communication policy.}
% \yawen{comment}
% 好像放到外面整个同位语或者冒号比较好。
% 这个图貌似还是柱状图比较好啊，一般表示分布的话是不是都是柱状图

% 这个图还是折线图or原来的柱状图比较好吧，就能看出来哪些边的通讯量高哪些低
% 这一段的写法太像描述方法或者实验了，如果单独写一个case study的章节，可能第一段这么写行，写在intro里不用这么写。就类似我们发现现有的通讯方法其通讯分布不平衡，如图所示的T2MAC和i2c，我们统计了每次任务执行中每个通讯链路的通讯量，即通讯次数（我们假定每次通讯的数据量相差不大），我们发现通讯分布是不平衡的。这种过于集中的xxx会导致xxx，例如一旦重要的通讯链路发生问题，整个系统会瘫痪
% \xuyan{comment}
%通信链路很多，所以横坐标的值的数量会很多，柱状图可能要从高到低排列，然后省略中间的一些内容

% This paper aims to develop a robust multi-agent communication policy learning method capable of defending various adversarial attacks.
{This paper aims to develop a learning method that enables multi-agent systems to adopt more robust communication policies, facilitating decentralized communication among agents without compromising task completion, so as to strengthen its ability to resist various attacks.}
% \rev{This paper aims to develop a learning method to enhance the robustness of multi-agent communication policies to ensure their performance in the face of various abnormal conditions. To this end, we borrow the decentralization theory and adjust the existing communication policy to the decentralized mode to improve its robustness.}

% \jie{comment}
%要不我们就不提two challenges了，感觉听起来也不是啥很严重的挑战。要不就改成类似这种。有些细节你们再对一下，可能不是很确切。
% \jie{question}
%这个方法训练得到的这个系统，是能抵御所有的攻击方法吗？例如那种对于observation的攻击那种？
Borrow from the idea of `fighting fire with fire', we propose {\tool}, which employs the adversarial training strategy to enable the training of more robust communication policies capable of resisting potential adversarial attacks. 
To achieve this, we construct an adversary (named {\tooladv}) {to dynamically generate adversarial samples against the communication policy $\mathit{CP}$.} %\rev{for the communication policy $\mathit{CP}$ of the target MAS. }
{\tooladv} is trained to dynamically identify and mask critical communication channels within $\mathit{CP}$ at each time step, thereby effectively disrupting message exchanges between agents. 
This disruption allows the system to explore potential alternative communication policies, making it less reliant on any single communication channel, thereby facilitating a transition towards decentralized communication.
% \jie{comment}
%onstruct an adversary (named DMAC Adv) based on a trained communication policy model CP of the target MAS 这里我感觉不是很清楚，为啥交 based on 。这个adversary为啥是based on的。感觉是说对于某个系统的当前communication policy model，是不是trained 似乎也不是很关键。不知道除了based on，是否有其他词

{The key to implementing this adversarial training lies in training an effective {\tooladv} that can dynamically identify critical communication channels. We first
% \rev{To accurately} 
model the agents and their interactions to facilitate the identification of these critical channels.
% \rev{for identifying the critical communication channels, }
We utilize observations of agents and status information
% , and data 
from neighboring agents as key features.
% \rev{ for identifying these critical channels}.
{Since masking critical channels can negatively impact the performance of the target MAS, we design a reward function for the adversary with dual objectives. 
It {maximizes the reduction in the reward of target MAS} %\rev{minimizes the reward for the target MAS} 
while minimizing the number of masked channels, preventing indiscriminate masking that could obscure the identification of essential communication channels.}
% \jie{comment}
%minimizes the reward for the target MAS 这个对吗，怎么感觉应该不太对，我感觉是maximize？
% \xuyan{comment}
%在训练adv的时候，需要让target mas的reward尽量小（意味negatively impact），才能说明攻击成功
% \jie{comment++}
%从这个说法上面，两个都是minimizes，感觉好像有点奇怪，不知道能不能稍微改个说法，换成一个是max 一个 min
% \xuyan{comment++}
%现在改成了最大化对reward的降低或者影响的程度
% \rev{Additionally, we design a reward function for the adversary, where masking critical channels negatively impacts the performance of the target MAS. This dual objective minimizes the reward for the target MAS while considering the number of masked channels, preventing indiscriminate masking that could obscure the identification of essential communication links.}
{With the trained {\tooladv} for generating adversarial samples against the target communication policy, we adopt adversarial training to retrain the communication policy.}
% \rev{Through this adversarial training process, the adversary generates adversarial samples for the target communication policy, prompting a retraining of the policy.} 
This encourages a reduced reliance on critical communication channels, facilitating a transition toward a decentralized structure and ultimately enhancing the system's robustness.

We evaluate {\tool} with two popular communication policies (T2MAC \cite{SunZLLXWZ24} and I2C \cite{DingHL20}) in four popular multi-agent tasks,
% \xuyan{comment} %还有一种说法是communication protocol
and compare it with two commonly-used and state-of-the art baselines. Results show that {\tool} can improve not only the robustness against various attacks but also the performance under normal situations. Specifically, in the face of learned adaptive attack, {\tool} improves the robustness of T2MAC by 47.9\%-81.9\% and I2C by 54.4\%-99.0\% on four tasks compared to the baselines. In the face of heuristic attack, {\tool} improves the robustness of T2MAC by 37.9\%-90.5\% and I2C by 38.3\%-117.7\% on four tasks compared to the baselines. Furthermore, we evaluate the performance of the communication policy under normal conditions, and {\tool} also shows optimal results. Finally, in order to evaluate whether {\tool} can achieve the decentralized communication policy, we carry out statistics on the communication frequency of each communication channel between agents, and the results show that {\tool} can make the communication policy more decentralized without increasing much communication cost. More evaluation results can be found in the appendix.
% \yawen{comment}
% {\tool} can achieve decentralization? 应该是前面的说法{\tool} can achieve the decentralized communication policy，或者如果觉得重复，可以是make the target communication policy more decenrailized。再检查一下类似随意的表达
% \xuyan{identifying and cutting off critical communication connections is a key step to achieve decentralization. Cutting off non-critical communication connections has a small impact on the behavior of agents, making it difficult to change the existing communication policy.} Therefore, we evaluate the ability of the component {\tooladv} to identify critical communication connections.
% \yawen{comment}
% communication frequency看看其他工作有没有类似的概念，如果没有可以根据fig 1的那个方式表述
% Compared to removing communication connections randomly or heuristically, {\tooladv} can both significantly reduce the performance of the communication policy, demonstrating that {\tooladv} can accurately identify critical communication connections.

In summary, this paper makes the following contributions:
\begin{itemize}[leftmargin=*]
    \item We apply decentralization theory to enhance the robustness of multi-agent communication policy.
    \item We introduce {\tool}, a learning method to enhance the robustness of multi-agent communication policy, which utilizes critical communication channels identification and adversarial learning.
    \item We conduct extensive evaluations on four categories of experimental environments and compare with two commonly-used and state-of-the art baselines, demonstrating promising results.
    \item We release the source code of {\tool} and detailed experiment results to facilitate the replication and further research\footnote{More details can be found on our website: \url{\website}}.
\end{itemize}
\section{Related Work}
\label{sec:rw}
\textbf{Communicative Multi-Agent Reinforcement Learning (CMARL).} 
There has been extensive research on encouraging communication between agents to improve performance on cooperative or competitive tasks. A communication policy defines how to decide whether to communicate with potential communicators for the purpose of message transportation \cite{ZhuD024}. Communication policies can be predefined or learned. Early communication policies used full-connection communication \cite{commnet,tarmac,maddpg}, where each pair of agents would be connected and transmit messages in a broadcast manner. Subsequent predefined partial communication \cite{JiangDHL20,ZhangZL19} causes each agent communicating with a limited number of agents, rather than all others. In recent years, learning to determine how to build communication structures between agents has provided the ability to generalize to more scenarios \cite{SunZLLXWZ24,MengT24}, and it has become increasingly popular because of its flexibility. Learnable communication policies is divided into individual control \cite{DingHL20,MaoZXGN20} and global control \cite{KimMHKLSY19,LiuWHHC020}. In individual control, each agent actively and individually decides whether to communicate with other agents. On the contrary, global control constructs a globally shared communication policy which can endow more precise control of communication links between agents.

\noindent
\textbf{Adversarial Attacks and Defenses in CMARL.} 
% There have been some researches on adversarial attack for multi-agent reinforcement learning, including modifying observation, action, communication, etc. Ilahi et al. \cite{IlahiUQJAHN22} considered attacks that rely on perturbing the state space, the reward space, the action space, and the model space, where one can perturb the model’s learned parameters.
Recently, the existence of adversarial communication in MARL has attracted increasing attention. 
For the adversarial attack of communication, much of the work has focused on directly attacking the victim by perturbing with the designated victim's observations or messages. Tu et al. \cite{TuWWMRU21} trained an attacker to learn how to generate adversarial perturbation and add them as noise to the victim agent's message. VSA \cite{MaL23} constructed an attacker which generates adversarial perturbations on the received communication messages by disturbing its controlled malicious agent to make non-optimal actions. These attack methods often select the victim communication channel randomly or based on rules, resulting in inefficiency or lack of flexibility.
As for the defense method, AME \cite{SunZHLFGH23} proposed a defense method by constructing a message-ensemble policy that aggregates multiple randomly ablated message sets. \cite{abs-2012-00508} adopted a Gaussian Process-based probabilistic model to compute the posterior probabilities that whether each partner is truthful to achieve robust communication.
To achieve robust communication, a Gaussian Process-based probabilistic model \cite{abs-2012-00508} was adopted to compute the posterior probabilities that whether each partner is truthful.
$\mathfrak{R}$-MACRL \cite{Xue0AROY22} learned an anomaly detector and a message reconstructor to recover the true messages, to maintain multi-agent coordination under message attacks. However, these methods often add extra processing to the abnormal messages without really adjusting the existing communication policies.
% However, generalizability is limited because they only consider defense against perturbations to message content and not against attacks on communication policies.

\section{Approach}
\label{sec:problem}

% \subsection{Problem Setting}
\subsection{Task Formulation}
We consider a problem setting where a MARL joint policy $\pi = \{\pi_1,...,\pi_n\}$ has been effectively trained for $n$ agents in the MAS, which determines the agent's actions. In addition, there is a communication policy model $\mathit{CP}$ which defines how to make the decisions of whether to communicate with potential communicatees to enable message transferring \cite{ZhuD024}.  %Because in the current multi-agent systems, the message is often delivered in the form of observation, so we use ``observation'' to represent the obtained message by the agent.} %相关文献中的说法
% defines whether two agents can communicate with each other.
% \yawen{comment}
%  Because 这句貌似没啥必要
% 现有的communication policy都是defines whether two agents can communicate with each other？这里需要高度概括且全面的一个表述
% 再就是CM跟pi的关系最好能定义出来，比如pi是包含CM，还是两者是并列的，然后合起来是个啥，或者参考一下communication policy的文章，看看有没有定义
The decision model $\pi$ of the agents usually utilizes the information passed through the communication policy $\mathit{CP}$ to determine the next action.
At each time step $t$ in an episode, the $i$-th agent receives a local observation $o_{i}^t$ from the global state $s_t$ according to $\mathit{CP} : s^t \rightarrow o_i^t$.
% \yawen{comment}
% 这句可以加一个类似低下pi的形式化表述，o经过CP产生一个输出
The individual policy $\pi_i$: $o_i \rightarrow a_i$ describes how the $i$-th agent chooses its action $a_{i}^t$ based on its local observation $o_{i}^t$. The joint action $a^t = \{a_{1}^t,...,a_{n}^t\}$ results in the next state $s_{t+1}$ with the state transition probability $P(s_{t+1}|s_t,a_t)$. Consequently, a global reward $r_t$ is obtained according to reward function $R(s_t,a_t,s_{t+1})$. In our problem, we focus on enhancing the robustness of communication policy $\mathit{CP}$.
% \xuyan{Therefore, our goal is to use the reward function as a guide for adversarial training of $\mathit{CP}$, thereby increasing its robustness to various attacks.}
% \yawen{comment}
% 为啥用Therefore，前后没因果关系啊
% 这段都是在介绍MARL以及通讯的问题设置，跟最后一句没啥关系吧
% \xuyan{comment}
%这段最后是否需要加一个我们的目标

\subsection{{\tool}}
\begin{figure}[tbp]
    \centering
    \includegraphics[width=0.5\textwidth]{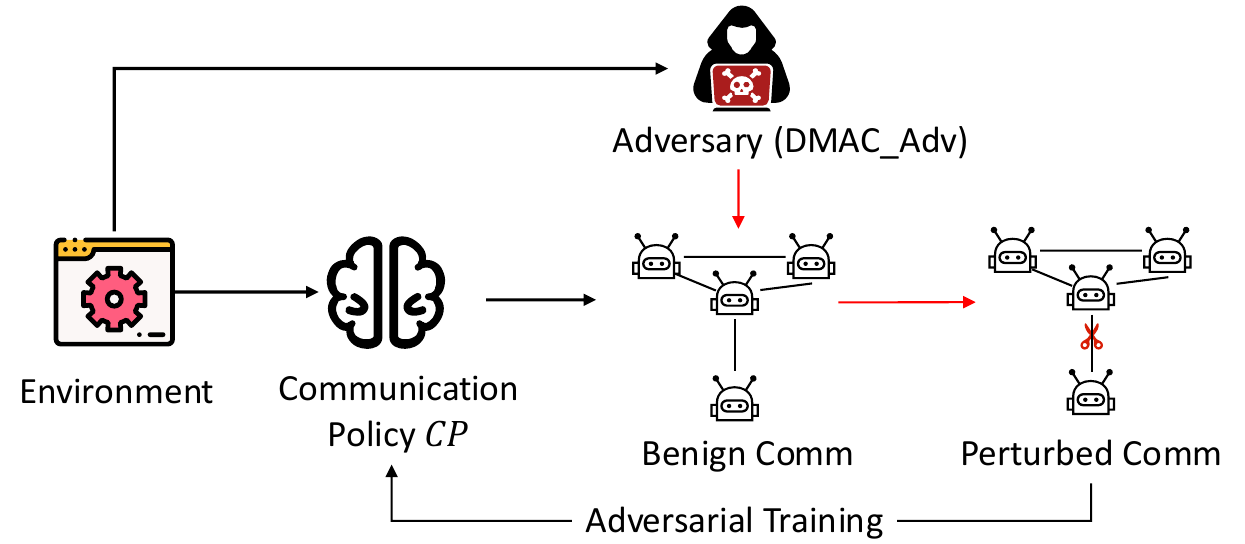}
    \caption{The overview of our proposed {\tool}.}
    \label{fig:overview}
% \vspace{-0.1in}
\end{figure}

We draw inspiration from the decentralization theory \cite{boss2023robustness} and aim to enhance the robustness of the communication policy by adjusting its relatively centralized pattern
% unbalanced pattern
% \jie{comment}
%我感觉从unbalanced pattern到decentralized pattern不是一个很对称的说法，感觉应该是说从相对集中的到分散式的，或者就换个其他表达。
to a decentralized pattern. 
Based on this, we propose {\tool}, as shown in Figure \ref{fig:overview}.
% \rev{which adjusts existing communication policy and ultimately enhances its robustness by identifying critical communication channels dynamically at each time step and achieving decentralization via adversarial training.}
% \jie{comment}
%你这里直接叫做 critical communication channels identification呢，感觉就很直接。你这个center 感觉碰到较真的，就会问这个center的一定是critical的吗，感觉没有必要
%by 1）identifying critical communication channels dynamically at each time step and 2）achieving decentralization via adversarial training。
%然后下面的题目就叫做  identifying critical communication channels，achieving decentralization via adversarial training 这样可能也比较确切
% Drawing on the concept of `fighting fire with fire', we implement decentralization of communication policies by adversarial learning.
There is an existing communication policy $\mathit{CP}$ being trained, which takes environmental information as input and outputs a benign communication decision.
% \yawen{comment}
% communication policy输出benign communication policy？
% {\tooladv}输出masking policy？
% We treat the well-trained model of center identification as an adversary, 
We construct an adversary (called {\tooladv}) to identify and mask the critical communication channels, which also takes environmental information as input and outputs adversarial samples (i.e., masking decision) to perturb the channels between certain agents. 
With the adversarial samples generated by {\tooladv}, the communication policy is retrained in the process of adversarial training.
In this way, critical communication channels in $\mathit{CP}$ %the existing communication policy 
are masked and $\mathit{CP}$ is forced to adjust to a decentralized mode.

As a key step in implementing the decentralized communication policy, we give detailed description of {\tooladv} in the following sections, including the problem modeling, feature extraction module and the architecture \& training of {\tooladv}.
% by implementing this adversary during the adversarial training process of $\mathit{CP}$.
% After the perturbation, the original unbalanced communication policy is destroyed and a decentralized perturbed communication policy is formed.
% \yawen{comment}

\begin{figure*}[tbp]
    \centering
    \includegraphics[width=\textwidth]{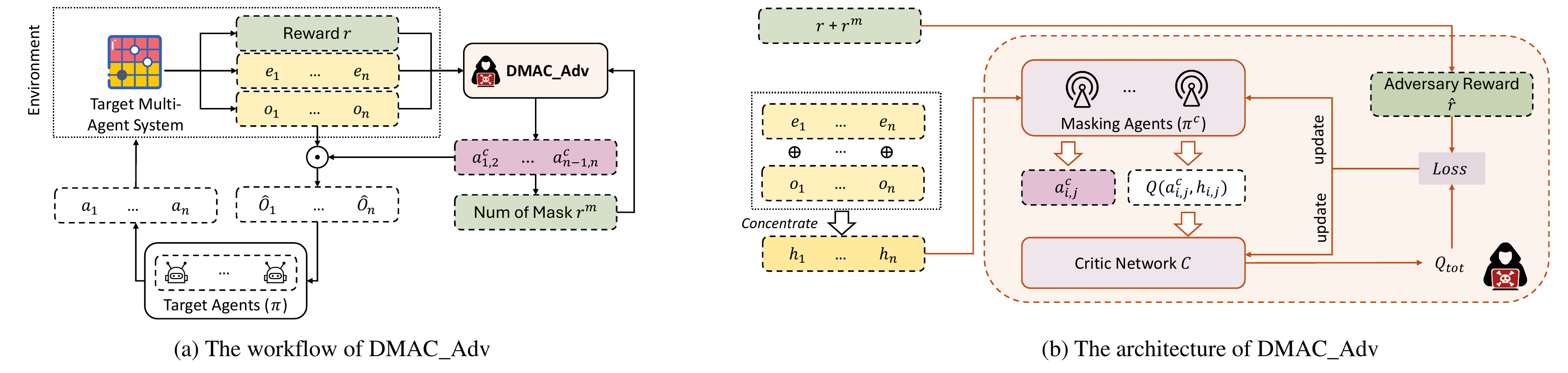}
    \caption{The overview of {\tooladv}. (a) At each time step, {\tooladv} obtains the feature of each agent from the multi-agent environment, and outputs a decision for each communication channel whether or not to mask. (b) During the training, the masking agents' policy network learns the masking action and individual value, and the critic network learns the total value to estimate the expected reward. The loss function is introduced to minimize the difference between the expected reward and actual reward.}
    \label{fig:adv}
% \vspace{-0.1in}
\end{figure*}

\subsubsection{Problem Modeling}
% \textbf{\textit{Problem Modeling}}
\label{subsubsec:center}

We model the problem of identifying critical communication channels as a MARL problem, where multiple agents (called \textit{masking agents}) learn the policy of masking channels.
For each masking agent, its goal is to develop a policy (denoted as $\pi_{i,j}^c$, where $c$ represents ``communication'') that determines whether to mask the communication channel between the agents $i,j$ in the target MAS at each time step.
% The optimization goal of the adversary's policy is to minimize the reward of the target agents and minimize the number of cut. We model the learning of this policy as a MARL problem to learn how to identify and mask critical communication channels. Specifically, we introduce {\tooladv}, which contains multiple masking agents. For each agent, our goal is to develop a policy (denoted as $\pi_{i,j}^c$, where $c$ represents ``communication'') that determines whether to mask the communication channel between the agents $i,j$ in the original MAS at each time step.
% \yawen{comment}
% 下面这句完全看不懂，the same environment是想表达什么，上面讲完了{\tooladv}的目标，然后不是直接说他要实现这个目标问题要怎么定义吗？
% \xuyan{In the training process of {\tooladv}, we use the same environment with {\tool} and the agents in the environment is called as target agents, which have a fixed joint policy $\pi$.}
% We treat the target agents, with a fixed joint policy $\pi$, as part of the environment. 
The decision processes of masking agents is modeled as decentralized partially observable Markov decision process, which can be defined by a tuple $G = <\mathcal{S}, \mathcal{A}^c, O, P, R^c, \mathcal{N}, \gamma>$. $\mathcal{S}$ is the global state space. $\mathcal{A}^c$ is our defined action space $\{0,1\}$ of each masking agent. The $a^c_{i,j} \in \mathcal{A}^c$ represents the masking action of the masking agent, where $i,j \in \{1,...,n\}$ and $i < j$. $a^c_{i,j} = 1$ denotes masking and $a^c_{i,j} = 0$ not masking the communication channel between target agents $i,j$. $O$ denotes the observation function and defines the input of the masking agents. $P$ is the state transition probability, $R^c$ is our designed reward function, $\mathcal{N}$ is the number of masking agents which is calculated by $\frac{n(n-1)}{2}$
% the same as the number of communication channels
and $\gamma$ is the discount factor applied to the rewards. We define the policy of masking agent as $\pi_{i,j}^c : h_{i,j} \rightarrow a_{i,j}^c$, where $h_{i,j}$ is the input of masking agent which is detailed in the following section.

% \yawen{comment}
% 这几段哪个是定义问题，哪个是训练过程的，哪个是优化目标的，它们之间的关系是什么？
Following the above, the workflow of {\tooladv} 
% \rev{training process of center identification module} 
is illustrated in Figure \ref{fig:adv}a. {\tooladv} trains the policy $\pi^c$ for masking agents to take masking actions at each time step. 
The initial observation of target agent $i$ is $o_i$ which can be represented as $o_i = \{o_{i,j} \mid j = 1, \dots, n \text{ and } j \neq i\}$, where $o_{i,j}$ denotes the message between target agents $i$ and $j$. The observation after perturbation (i.e., mask the critical communication channel) is defined by the following operation:
\begin{equation}
\label{equ:obs}
\hat{o}_{i,j} = o_{i,j} \odot a_{i,j}^c = 
\begin{cases}
o_{i,j}, & \text{if} \; a_{i,j}^c = 0,\\
null,  & \text{if} \; a_{i,j}^c = 1,
\end{cases}
\end{equation}
where $a_{i,j}^c = 0$ indicates that the communication channel between target agents $i,j$ remains normal. Otherwise, when $ a_{i,j}^c = 1$, the communication channel is masked and closed.
% \yawen{comment}
% 这句没懂，i>j是不是就是跟i<j一样？这句话感觉可以换种表达
% \xuyan{Note that, in Equation \ref{equ:obs}, we assume $i < j$ for the sake of representation. When $i > j$, we can use $a_{j,i}^c$ to get the new observation of agent $j$.}
{Note that, %for reducing overhead, 
both target agents $i$ and $j$ utilize $a_{i,j}^c$ ($i < j$) to determine the final observation.}
% when $i > j$, we use its symmetric terms $a_{j,i}^c$ to determine the observation.}
The final observation of target agent $i$ is $\hat{o}_i = \{ \hat{o}_{i,j} \mid j = 1, \dots, n \text{ and } j \neq i \}$. Based on the new observations $\{\hat{o}_1, \dots \hat{o}_n\}$ obtained from $\mathit{CP}$, each target agent selects an action through its fixed policy $\pi$.
% \yawen{comment}
% 所以是its fixed policy $\pi$还是通讯策略？
The actions $\{a_i, \dots, a_n\}$ act on the environment to move on to the next state. 
% \yawen{comment}
% % 上面的target agent指的是cp还是pi？可能要从CP的角度说？
% \xuyan{comment}
%上面的target agent指的是目标系统里智能体，与cp和pi无关，就是我们使用的环境里的agent

The objective of $\pi_c$ is to maximum the total reward of masking agents $\hat{r}$, which is calculated by the reward of the target MAS $r$ and the number of mask $r^m$, and $\hat{r}$ is inversely related to them:
\begin{equation}
\label{equ:obj}
    \begin{aligned}
    & obj(\pi_c) = \arg \max \hat{r} = \\
    &  \arg \max \frac{1}{\sum_{t=0}^{T} \gamma^t (w_1 \cdot R(s_t,a_t,s_{t+1}) + w_2 \cdot R^m(a^c_t)) + \xi} 
    \end{aligned}
\end{equation}
where $R^m(a^c_t) = \sum_{i=1}^{n-1} \sum_{j=i+1}^{n} a_{i,j}^{c}$, which counts the number of masked communication channels at each time step $t$, 
$w_1$ and $w_2$ are two hyperparameters to adjust the weights of the two factors, $\xi$ is a small value to avoid zero denominator.

\subsubsection{Feature Extraction}
\label{subsec:input}
% This part details the input of the adversary. 
Using observations to determine communication policy is a common setting, however, this setting lacks a description of the relationships between agents and the team structure \cite{MaWWXWLX024}, making it difficult to mine critical communication channels.
Therefore, we model the agents and their interactions where we not only use the observations but also status information from neighboring agents as features for identifying the critical communication channels.
% To better represent the status of each agent and the relationship between agents, we
% the relationship between agents and their neighbors. 
% \yawen{comment}
% 这块为了更好表示所做的设计，前面也没有任何铺垫，或是能对应上某个挑战？

We obtain the observation of the each agent $\{o_1,...,o_n\}$ from the environment and the representation of each agent $\{e_1,...,e_n\}$ (called \textit{embedding representation}) based on the agent itself and its neighbors.

% As described in Section \ref{subsec:model}, the target agents team can be modeled as an undirected graph. 
Specifically, we model the target agents team and interactions between teammates as an undirected graph, with each agent represented as a vertex. The state information and cooperative relationships of the target agents are effectively represented by the properties of the vertices, and the weights of the edges. Each vertex holds a set of properties for the corresponding agent, such as, location, speed, health, and so on. The edges between vertices and their weights are the interactions and situations between two agents. The weight of an edge is determined by the distance between two agents, as obtained from the environment, indicating the degree to which the two agents can interact with each other. 
Note that, the distance is calculated differently depending on the environment. For example, in StarCraft II \cite{SamvelyanRWFNRH19}, the distance is calculated by the euclidean distance between the coordinates of two soldiers. In Traffic Junction \cite{SinghJS19}, the distance represents the number of grids between two cars.
% \yawen{comment}
% 不同环境都是用的distance吗
% As we can see in Figure \ref{fig:adv}a, {\tooladv} can obtain the observation of the each agent $\{o_1,...,o_n\}$ and the embedding representation of each agent $\{e_1,...,e_n\}$ from the environment.
% \yawen{comment}
% embedding representation指的是图表示的结果吗？前面也没铺垫突然出来这个词很迷惑
% 另外考不考虑增加实验证明这个设计的必要性，可以加到appendix
The embedding representation is calculated as follows:

% \xuyan{comment} %直接复制了issta的内容，需要换说法
{\tooladv} firstly initializes the embedding representation of each vertex based on its own attribute. Then, {\tooladv} aggregates the information from neighboring vertices using a weight-based aggregation function. 
For a vertex $v$, $\mathcal{N}(v)$ is the neighbor vertices of $v$ and $e^{k}_{v}$ is the embedding representation of vertex $v$ after the $k$-th iteration.
$w(u,v)$ is the weight of the edge between vertices $u$ and $v$, quantifying the interaction between the two agents and how they interact. We represent the weight by the distance between two agents, obtained from the environment, which indicates the extent to which the agents can interact.
Then the embedding of $v$ at $(k+1)$-th iteration $e^{k+1}_{v}$ is the sum of $e^{k}_{v}$ and the weighted and averaged neighbor embedding, which is specified as follows:
\begin{equation}
\label{equ:aggregate}
    e^{k+1}_{v} \leftarrow Sum(e^{k}_{v}, Mean({\frac{1}{w(u,v)} * e^{k}_{u} ), u \in \mathcal{N}(v) }).
\end{equation}
In this way, we can obtain the embedding representation of each vertex as $e_{v}$.

As shown in Figure \ref{fig:adv}b, we concatenate the observation and embedding representation of each agent to get the feature vector $\{h_1,...,h_n\}$, where $h_i = o_i \oplus e_i$. 
% \yawen{comment}
% \odot前面用过了，有别的含义了
Then, we send $h_i$ and $h_j$ to the masking agent for subsequent decision making about whether to mask the communication channel between target agents $i$ and $j$. 
% Then, we merge the representation of two agents $i,j$ and input the feature into the BiLSTM \cite{ZhangZHY15} model:
% \begin{equation}
% \label{equ:lstm}
%     h_{i,j} = BiLSTM(g_i, g_j).
% \end{equation}
% BiLSTM can effectively eliminate the effects of different order of inputs (agents). Based on this, we obtain the agent-pair embedding $h_{i,j}$ for agent $i$ and agent $j$.
% \yawen{comment}
% merge是个啥操作。为啥BiLSTM可以消除影响。实际上我们是对每个i和j对进行了表示，然后masking agent拿这个做输入，输出这个对是否断开？

\subsubsection{The Architecture and Training of {\tooladv}}
{\tooladv} consists of two networks: the policy network of masking agents $\pi^c$ and the critic network $C$. Specifically, the network $\pi^c$ learns the policy of masking agents, taking $[h_{i},h_{j}]$ as input and outputting the masking actions $a^c_{i,j}$, where $i$ and $j$ represent two agents. For convenience, we use $h_{i,j}$ instead of $[h_{i},h_{j}]$ in the following paper. The critic network $C$ is designed to evaluate the joint actions of all masking agents from a global perspective.

Following value-based CTDE, we evaluate the value of masking actions while learning $\pi^c$. First, we train $\pi^c$ to learn individual value function $Q(a^c_{i,j}, h_{i,j})$ for each masking agent, which assesses the individual policy. This value function represents the benefit of taking action $a^c_{i,j}$ given the feature vector $h_{i,j}$. Second, $C$ learns the centralized value function $Q_{tot}(a^c_{i,j}, h_{i,j})$ to evaluate the collective policy, providing an estimate of the adversary reward $\hat{r}$.

To maximize both individual and total values simultaneously, this can be achieved by adhering to the Individual-Global-MAX (IGM) principle \cite{HongJT22}:

% \begin{equation}
% \label{equ:igm}
% \begin{aligned}
%     & \arg \max_{a^c \in \mathcal{A}^c} Q_{tot} (h, a^c)= \\
%     & \left<\arg \max_{a^c_{1,2}} Q(h_{1,2}, a^c_{1,2}), \dots, \arg \max_{a^c_{n-1,n}} Q(h_{n-1,n}, a^c_{n-1,n}) \right>
% \end{aligned}
% \end{equation}

\begin{equation}
\label{equ:igm}
\arg \max_{a^c \in \mathcal{A}^c} Q_{tot} (h, a^c)= \left( 
\begin{aligned}
  &\arg \max_{a^c_{1,2}} Q(h_{1,2}, a^c_{1,2}) \\
  &\dots \\
  &\arg \max_{a^c_{n-1,n}} Q(h_{n-1,n}, a^c_{n-1,n})
\end{aligned}
\right)
\end{equation}

% \begin{equation}
%     \label{equation:igm}
%     \begin{aligned}
%     & \arg\max_{a^m\in \mathcal{A}^{m}} Q_{tot}(o, a^m; \omega) = \\
%     & \left< \arg\max_{a^m_1} Q_1(o_1, a^m_1; \theta), ..., \arg\max_{a^m_n} Q_n(o_n, a^m_n; \theta) \right>.
%     \end{aligned}
% \end{equation}

Specifically, in {\tooladv}, we constrain the weights $\omega$ of the critic network $C$ to be non-negative to ensure adherence to the Equation \ref{equ:igm}, which can be defined as follows:
\begin{equation}
\label{equ:weights}
\begin{aligned}
    &Q_{tot}(h, a^c; \omega) = \omega (Q(h_{i,j}, a^c_{i,j})), \forall{i,j} \in \{1, \dots, n\}, \\ 
    &i \neq j; \omega \geq 0.
\end{aligned}
\end{equation}
Equation \ref{equ:weights} can be replaced with other implements such as VDN \cite{SunehagLGCZJLSL18}, OPLEX \cite{WangRLYZ21}, and QTRAN \cite{SonKKHY19}. 
% \xuyan{comment} %必要时可删除

Finally, we use TD losses \cite{HausknechtS15} as the loss function of {\tooladv} for iterative optimization of $\pi^c$ and $C$, to minimize the error gap between the expected value and the estimate value $Q_{tot}(h, a^c)$.
\begin{equation}
    \mathcal{L} = \arg \min \mathbb{E} [(y_{tot}-Q_{tot}(h,a^c))^2]
\end{equation}
where $y_{tot} = \hat{r} + \gamma \max_{\Tilde{a}^c} \Tilde{Q}_{tot} (\Tilde{h}, \Tilde{a}^c)$, which denotes the expected value. $\hat{r}$ is the adversary reward introduced in Equation \ref{equ:obj} and $\Tilde{Q}_{tot} (\Tilde{h}, \Tilde{a}^c)$ is calculated for the next time step by the stale network. Since Equation \ref{equ:igm} holds, we can perform the maximization of $Q_{tot}$. Previous work \cite{LiuZC23} has demonstrated the feasibility and stability of implementing updates in this manner.
Algorithm \ref{alg:adv} briefly presents the training process of {\tooladv}.

\begin{algorithm}
\caption{The training algorithm of {\tooladv}}
\label{alg:adv}
\begin{algorithmic}[1]
\State \textbf{Input:} The policy $\pi$ of target agents, the observation $\{o_1, \dots, o_n\}$ and the feature vector $\{h_{1}, \dots, h_{n}\}$ of each agent.
\State \textbf{Output:} The policy $\pi^c$ of masking agents
\State \textbf{Initialization:} The networks of $\pi^c$ and $C$ 
% \State $x \gets 0$  \Comment{Initialize sum t}
\For{$each$ $training$ $batch$}
    % \State $x \gets x + a_i$
    \State Get joint masking action and values from $\pi^c$:
    \State $\{a_{1,2}^c, \dots, \}, \{Q(h_{1,2},a_{1,2}^c), \dots, \} = \pi^c(h_{1,2}, \dots, )$
    \State Get the number of masking $r^c$
    \State Get the final observations:
    \State $\{\hat{o}_1, \dots, \hat{o}_n\} = \{o_1, \dots, o_n\} \odot \{a_{1,2}^c, \dots, a_{n-1,n}^c\} $
    \State Get joint action from $\pi$
    \State $\{a_1, \dots, a_n\} = \pi(\hat{o}_1, \dots, \hat{o}_n)$
    \State Execute joint action of target agents and get reward $r$ from environment
    \State Calculate the global value $Q_{tot}$ using network $C$
    \State Update $\pi_c$ and $C$ by the TD loss with $r$, $r^c$ and $Q_{tot}$
\EndFor
\State \textbf{return} $\pi_c, C$
\end{algorithmic}
\end{algorithm}

% \input{fig_tex/fig_overview}

% \subsubsection{Decentralization Implementation}
% \yawen{comment}
% 第二个不应该是利用adversary进行对抗训练之类的吗?
% Drawing on the concept of `fighting fire with fire', we implement decentralization of communication policies by adversarial learning.
% As shown in Figure \ref{fig:overview}, there is an existing communication policy model $\mathit{CP}$ being trained, which takes environmental information as input and outputs a benign communication decision.
% % \yawen{comment}
% % communication policy输出benign communication policy？
% % {\tooladv}输出masking policy？
% {We treat the well-trained model of center identification as an adversary, which also takes environmental information as input and outputs adversarial samples (i.e., masking decision) to perturb communication channels between certain agents. In this way, critical communication channels in the existing communication policy are masked and $\mathit{CP}$ is forced to adjust to a decentralized mode by implementing this adversary during the adversarial training process of $\mathit{CP}$.}
% % After the perturbation, the original unbalanced communication policy is destroyed and a decentralized perturbed communication policy is formed.
% % \yawen{comment}

% \input{sec/4_approach}
\section{Experiments}
\label{sec:experiment}

\subsection{Experimental Setup}
\label{subsec:setup}

\subsubsection{Research Questions}
We consider the following research questions:
% \noindent
% 1) Can {\tool} improve the \textbf{robustness} of communication model?

% \noindent
% 2) Can {\tool} improve the \textbf{performance} of communication model?

% \noindent
% 3) Can {\tool} implement a \textbf{decentralized} communication policy?
\begin{itemize}
    \item Can {\tool} improve the \textbf{robustness} of communication policy?
    \item Can {\tool} improve the \textbf{performance} of communication policy?
    \item Can {\tool} implement a \textbf{decentralized} communication policy?
    % \item Can {\tool} accurately identify critical communication connections?
    % \item Does {\tool} help average communication policy? 
\end{itemize}

\begin{table*}[h]
\renewcommand\arraystretch{1.1}
\centering
\caption{The win rate of the target MAS with two communication policies under attack before and after robustness optimization. }
\label{result:robust}
\resizebox{0.8\textwidth}{!}{
\begin{tabular}{c|cccc|cccc} \toprule 
\textbf{Attack Method} & \multicolumn{4}{c|}{\textbf{Learned Adaptive Attack}} & \multicolumn{4}{c}{\textbf{Heuristic Attack}} \\ \midrule
 \diagbox{\textbf{Method}}{\textbf{Env}}& \textbf{SC} &  \textbf{CN} &\textbf{TJ} & \textbf{PP}  & \textbf{SC} &  \textbf{CN} &\textbf{TJ} & \textbf{PP} \\ \midrule 
 \textbf{T2MAC}& 27.8\%   & 30.1\%   & 45.8\%   & 37.9\%   & 32.1\%  & 37.7\% & 43.6\% & 39.1\% \\
 \textbf{+AME}   & 37.6\%   & 42.4\%   & 50.8\%   & 40.5\%   & 39.3\%  & 48.2\% & 53.5\% & 43.7\% \\
 \textbf{+$\mathfrak{R}$-MACRL} & 33.2\%  & 39.8\% & 47.4\% & 38.2\% & 36.7\% & 42.6\% & 49.1\% & 40.2\% \\
\textbf{+{\tool}}  & \textbf{60.4\%} & \textbf{62.7\%} & \textbf{70.4\%} & \textbf{66.3\%} & \textbf{69.9\%} & \textbf{70.1\%} & \textbf{73.8\%} & \textbf{74.4\%} \\
 \hline 
\hline

 \textbf{I2C}& 22.4\% & 24.3\% & 37.7\% & 32.5\% & 27.4\% & 30.8\% & 38.2\% & 35.7\% \\
\textbf{+AME}   & 33.9\% & 38.1\% & 43.6\% & 37.2\% & 34.4\% & 46.1\% & 50.7\% & 40.6\% \\
\textbf{+$\mathfrak{R}$-MACRL} & 29.6\%  & 30.1\% & 40.2\% & 35.9\% & 30.5\% & 39.4\% & 42.7\% & 38.2\% \\
\textbf{+{\tool}} & \textbf{58.3\%} & \textbf{59.9\%} & \textbf{67.3\%} & \textbf{62.5\%} & \textbf{66.4\%} & \textbf{68.2\%} & \textbf{70.1\%} & \textbf{71.9\%} \\
 \bottomrule
\end{tabular}}
\end{table*}

\subsubsection{Multi-Agent Environments}
Our experiments are conducted on four popular multi-agent benchmarks with different characteristics.

\textbf{StarCraft Multi-Agent Challenge (SC)}. SMAC \cite{SamvelyanRWFNRH19} is derived from the well-known real-time strategy game StarCraft II. It delves into micromanagement challenges where each unit is steered by an independent agent making decisions under partial observation. 
\begin{figure}[htbp]
    \centering
    \includegraphics[width=0.4\textwidth]{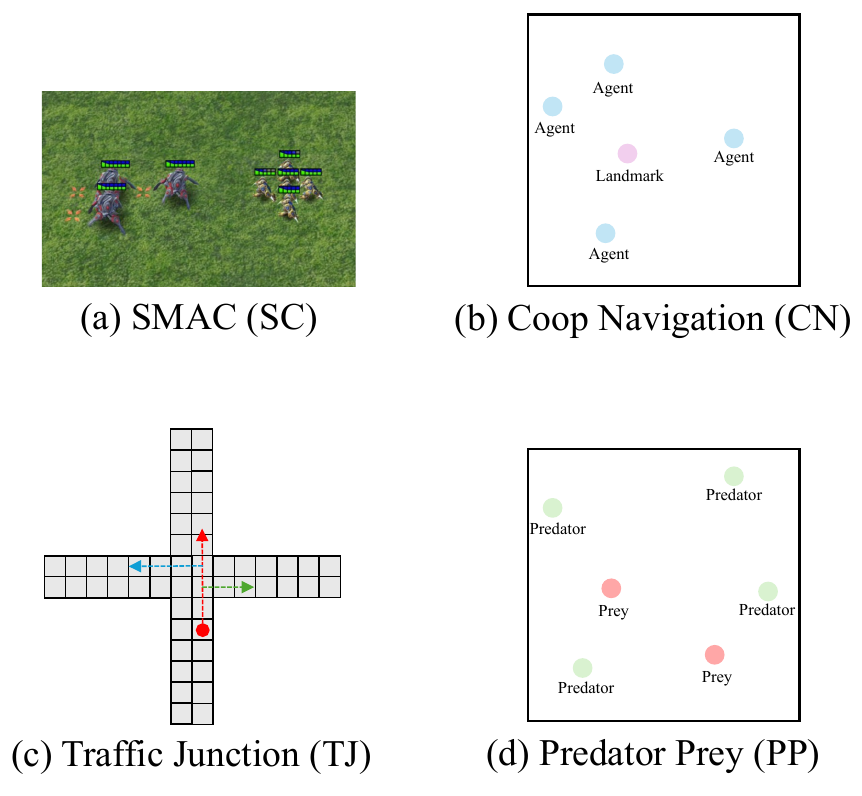}
    \caption{Multiple environments used in our experiments.%\xuyan{comment} %pp画错了,predator和prey标反了
    }
    \label{fig:env2}
% \vspace{-0.1in}
\end{figure}

% \begin{figure*}[htbp]
%     \centering
%     \includegraphics[width=\textwidth]{fig/env.pdf}
%     \caption{Multiple environments considered in our experiments.}
%     \label{fig:env}
% % \vspace{-0.1in}
% \end{figure*}
\noindent
We select a representative map $1c3s5z$ as the experimental map, which contains 9 agents in three categories.

Multi-Agent Particle Environments \cite{MPE} is a vital MARL benchmark set in a 2D grid. We focus on two tasks, \textbf{Cooperative Navigation (CN)} and \textbf{Predator Prey (PP)}. In CN, the task for agents is to navigate to different landmarks, whereas, in PP, their objective is to capture unpredictably moving prey. CN contains 7 target agents and PP contains 8 target agents.

\textbf{Traffic Junction (TJ)}. The simulated traffic junction environment \cite{SinghJS19} consists of cars moving along pre-assigned potentially interesting routes on one or more road junctions. The goal of each car is to pass the road without collision. Following the same settings in previous work \cite{LiuWHHC020}, the number of target agents is 10. The action space for each car is \textit{gas} and \textit{break}, and the reward consists of a linear time penalty.

\subsubsection{Target Model}
Our experiments are based on existing learnable communication policies and enhancing their robustness. We choose the following two communication policy algorithms:

\textbf{T2MAC} \cite{SunZLLXWZ24} is a straight-forward and effective method that enables agents to learn selective engagement and evidence-driven integration.

\textbf{I2C} \cite{DingHL20} infers the impact of one agent on another and quantifies the necessity of agent-to-agent communication.

\subsubsection{Evaluation Metrics}
We use task completion to evaluate the performance and robustness of the communication policy.
For different tasks, the evaluation criteria of completing task are slightly different. For example, for SC, defeating the opponent means completing the task, and for CN, completing the task requires reaching the destination without collision. Therefore, for the sake of simplicity, in the following paper, we uniformly use \textbf{win rate} to carry out experimental evaluation, that is, task completion rate.
Additionally, we report the win rate under the following two types of attack methods following previous works \cite{SunZHLFGH23}:

\noindent
(1) \textit{Learned adaptive attack} learns the strongest adversarial communication with an RL algorithm to minimize the victim’s reward (a white-box attacker which knows the victim’s reward). The learned attack can strategically mislead the victim agents. 
% As shown in prior works, the theoretically optimal attack (which minimizes the victim’s reward) can be formed as an RL problem and learned by RL algorithms.

\noindent
(2) \textit{Heuristic attack} perturbs messages based on heuristics. We consider randomly generated messages within the valid range of communication messages.

% In addition, we use \textbf{attack ratio (\%Attack)} to measure the stealth of attack methods, which is calculated as follows: There are n target agents, so there are $\frac{n(n-1)}{2}$ potential communication connections. There are $T$ time steps in a task. If the number of communication connections attacked in each time step is $att_t$, the attack ratio is calculated by $\frac{\sum_{t=1}^{T} att_t}{T \cdot \frac{n(n-1)}{2}}$.

\subsubsection{Baselines}
To evaluate performance and robustness, we choose the following methods as our baselines:

\textbf{AME} \cite{SunZHLFGH23} is a certifiable defense by constructing a message-ensemble policy that aggregates multiple randomly ablated message sets.

\textbf{$\mathfrak{R}$-MACRL} \cite{Xue0AROY22} is a defense approach based on message refactoring, by constructing a two-stage message filter. It can identify malicious messages and recover them, in order to maintain multi-agent coordination under message attacks.

% \xuyan{comment} %以下baseline放到附录
% To evaluate the ability to identify critical communication connections, we used the following methods as baselines:

% \noindent
% (1) \textit{Random Attack} randomly select and cut off the communication connections without any guidance.

% \noindent
% (2) \textit{Reward-Based Attack} select and cut off the communication connections based on reward. Specifically, we use the reward of each agent as the basis, select the agent with the highest reward, and randomly select an agent from its neighbors, and cut off the communication connection between the two agents.
% Note that, Heuristic-Identify is different from heuristic attack in the evaluation metrics.
% \yawen{comment}
% note这句啥意思，不都是win rate，只不过一个越大越好 一个越小越好？

More experimental details can be found in the appendix.

\subsection{Results}

\noindent
\textbf{Robustness.} As Table \ref{result:robust} shows, we begin our robustness evaluation by comparing the win rate of target MAS with the communication policies enhanced by different methods in various environments.
% Now, we focus on the robustness of these communication models against attacks. 
% The results of robustness are shown in Table \ref{result:robust}, where we give the win rates of communication policies \rev{optimized with} \xuyan{enhanced by} {\tool} and other baselines under different types of attacks in four environments. 
% \yawen{comment}
% 上面两句合并一下
% 不太想用optimized这种说法，但是没想好怎么处理
% The details of learned adaptive attack and heuristic attack are put in \textcolor{blue}{Appendix}.
As we can see, when attacked, the win rate of the communication policy which has not been optimized decreases rapidly, while optimizing it with {\tool} can improve its robustness.
% \yawen{comment}
% % 要不要把算法origin的性能放到第一行？
% \xuyan{comment}
% Table1吗？相当于是没有面临攻击时候的性能也展示在表中？那可以考虑加一个阴影来明显的区分一下
For example, in environment SC, the original win rate of T2MAC is 81.2\% (which can be found in Table \ref{tab:rq2}), however, its win rate drops to 27.8\% after the learned adaptive attack. When optimized for T2MAC with {\tool}, it has a 60.4\% win rate against the same attack.
In addition, {\tool} also shows optimal results compared to other baselines in terms of improving the robustness of communication policies. 
% \yawen{comment}
% 在其他environment也optimal results？

\begin{table}[htbp]
\centering
\renewcommand\arraystretch{1.2}
\caption{The win rate of target MAS with two communication policies under normal conditions before and after robustness optimization. 
% ``Original'' represents the performance of the communication model without optimization.
% \yawen{comment}
% 这个original是不是你就选了一个T2MAC作为例子，是不是T2MAC，I2C都列上啊，要不跟Table 1一样？
}
\label{tab:rq2}
\resizebox{0.45\textwidth}{!}{
\begin{tabular}{c|cccc} \toprule 
 % & \multicolumn{4}{c|}{\textbf{T2MAC}} & \multicolumn{4}{c}{\textbf{I2C}}  \\ \midrule 
\diagbox{Method}{Env}   & \textbf{SC} &  \textbf{CN} &\textbf{TJ} & \textbf{PP}  \\ \midrule 
\textbf{T2MAC}     & 81.2\%  & 87.4\% & 91.7\% & 88.6\%  \\
\textbf{+AME}  & 82.2\%  & 88.3\% & 91.9\% & 89.2\%  \\
\textbf{+$\mathfrak{R}$-MACRL} & 80.9\%  & 84.9\% & 90.8\% & 88.0\%  \\

\textbf{+{\tool}} & \textbf{83.7\%} & \textbf{89.5\%} & \textbf{93.8\%} & \textbf{90.7\%}  \\  \hline 
\hline
\textbf{I2C} & 76.9\% & 80.1\% & 85.2\% & 82.4\% \\
\textbf{+AME}  & 77.1\% & 81.4\% & 87.3\% & 84.2\%  \\
\textbf{+$\mathfrak{R}$-MACRL} & 76.1\% & 79.8\% & 86.2\% & 82.9\% \\
\textbf{+{\tool}} & \textbf{79.4\%} & \textbf{83.6\%} & \textbf{90.4\%} & \textbf{88.1\%} \\

 \bottomrule
\end{tabular}}
\end{table}

\begin{figure*}[htbp]
    \centering
    \includegraphics[width=0.98\textwidth]{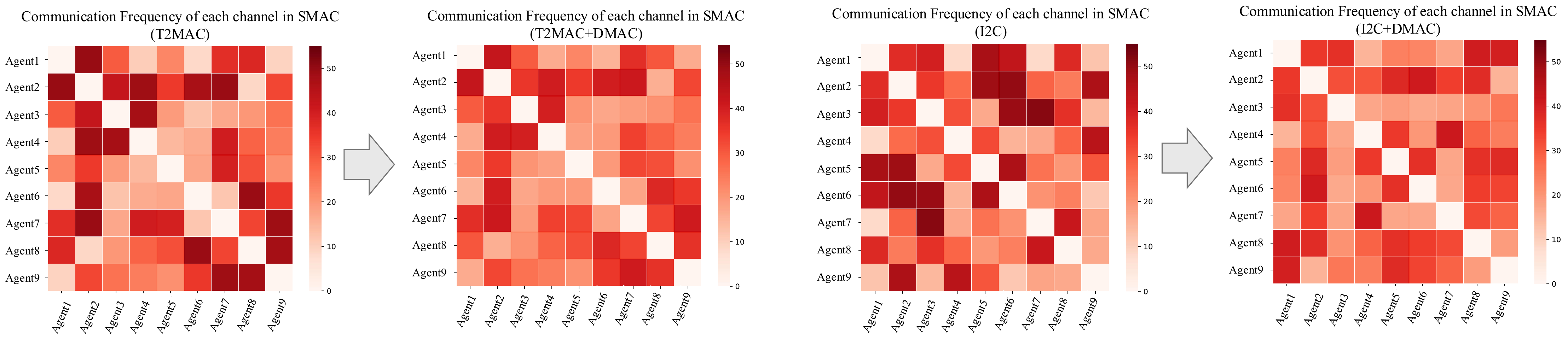}
    \caption{The communication frequency of the communication channel between each pair of agents in two environments. The darker the color, the more times the two agents communicate, that is, the higher the communication frequency of the communication channel. The grids on the diagonal represent the same agent and are therefore white.}
    % An example of test-time communication attacks in a communicative MARL system. (a) During training, agents are trained collaboratively in a safe environment, such as a simulated environment. (b) In deployment, agents execute pre-trained policies in the real world, where malicious attackers may attack and cut off the connections to mislead some victim agent(s).
    \label{fig:rq3}
% \vspace{-0.1in}
\end{figure*}

\noindent
\textbf{Performance.} Now, we focus on the performance of these communication policies under normal situation.
% We begin our evaluation by comparing the win rate of {\tool} with various baselines across various environments to test its overarching performance. 
From Table \ref{tab:rq2}, we can see that although the communication policies have shown good performance in various environments, {\tool} can still improve their performance. 
% \xuyan{comment} % good performance用在这里会不会有些主观
Compared with other baselines, the communication policy optimized by {\tool} has the highest win rate. For example, in SC, the original win rate of T2MAC is 81.2\% and the win rate becomes 82.2\% under the effect of AME. While $\mathfrak{R}$-MACRL makes the win rate drop to 80.9\% because it incorrectly modifies the correct message.
% \xuyan{comment} %写R-MACRL导致性能降低的原因
With {\tool}, the win rate of T2MAC increases to 83.7\%, which indicates the best performance compared with baselines. 
% Since the existing communication policies already perform well under normal conditions, our approach and baseline do not improve their performance much.

\begin{table}[bp]
\centering
\renewcommand\arraystretch{1.2}
\caption{The highest value (High), lowest value (Low), average value (Average), standard deviation (SD) of communication frequency among the communication channels. 
}
\label{tab:rq3}
\resizebox{0.48\textwidth}{!}{
\begin{tabular}{c|cccc|cccc} \toprule 
 % & \multicolumn{4}{c|}{\textbf{T2MAC}} & \multicolumn{4}{c}{\textbf{I2C}}  \\ \midrule 
\multirow{2}{*}{\diagbox{Method}{Env}}   & \multicolumn{4}{c|}{\textbf{SC}} &  \multicolumn{4}{c}{\textbf{PP}} \\ 
\cmidrule(r){2-5} \cmidrule(r){6-9} 
 & High & Low & Average & SD & High & Low & Average & SD \\ \midrule 
\textbf{T2MAC}     & 47.3  & 8.2 & 28.6 & 14.0  & 44.2  & 8.5 & \textbf{23.9} &12.6\\
\textbf{+{\tool}} & 42.5  & 15.2 & \textbf{28.4} & \textbf{9.0} & 36.5  & 15.5 & 26.1 & \textbf{7.2} \\  \hline 
\hline
\textbf{I2C} & 51.2  & 8.0 & 30.0&13.6 & 44.3  & 8.1 & \textbf{24.0} & 12.4\\
\textbf{+{\tool}} & 41.2  & 15.0 & \textbf{28.2} & \textbf{8.8} & 36.4  & 15.1 & 26.2 & \textbf{7.3}\\

 \bottomrule
\end{tabular}}
\end{table}

% \yawen{comment}
% 下面这段应该还没改过吧？有些结果如果要放到appendix，那就简单说一下appendix有验证xxx的实验就行了，相关结果分析不用在这展开，而是放到appendix，这里只写有结果的分析
\noindent
\textbf{Decentralization.} {As introduced in Section \ref{sec:intro}, the existing communication policies present unbalanced communication structure.
%, that is, a few edges communicate many times, and the communication utilization of most edges is very low. 
Then, we evaluate the situation of the communication policy after combining with {\tool}. 
% As shown in Figure \ref{fig:rq3}, the adjusted communication policy narrows the gap between the communication utilization of different edges, which means the communication connection is not concentrated in a few edges, but presents a decentralized structure. 
As Figure \ref{fig:rq3} shows, compared with before adjustment, the color of the grids in the communication frequency thermal map generated by the adjusted communication policy is more uniform. The previously dark-colored grids become lighter, while the previously light-colored grids become darker. This indicates that the communication frequency difference between communication channels is reduced in the adjusted communication policy.
Additionally, by comparing the average shown in Table \ref{tab:rq3}, we can see that {\tool} does not increase the communication cost much or even reduces the cost. 
Finally, we quantitatively analyze the degree of differentiation between these communication channels by calculating the standard deviation of their communication frequency. The higher the standard deviation means that the greater the difference in the communication frequency of the communication channel, that is, the more unbalanced. We can see that {\tool} reduces the gap between communication channels in communication frequency and achieves a more decentralized communication structure.}
% which is related with the communication frequency.}

\noindent
\textbf{More Results.} More evaluation results can be found in the appendix, including the accuracy of critical communication channel identification and ablation study.
\section{Conclusions}
\label{sec:conclusions}
Multi-agent communication plays an important role in multi-agent cooperative tasks, but when encountering abnormal situations, the non-robust communication policy often leads to the serious failure of the whole multi-agent system.
% communication policies are often vulnerable to adversarial attacks, which leads to serious failures.
{This paper proposes a multi-agent communication policy learning method {\tool}, which aims to enhance the robustness of the policy against various abnormal conditions. 
Inspired by decentralization theory, we achieve the goal of improving robustness by adapting existing communication policies to a decentralized structure. Technically, we firstly construct an adversary to identify and mask the critical communication channels. 
The policy learning of the adversary is modeled as a MARL problem, with dual optimization objective to minimize the reward of target MAS while considering the number of masked channels.
% difference between the expected reward and actual reward, where the reward function has been reconstructed. 
Then, we use the adversarial samples generated by the adversary to retrain the communication policy. In this way, the target communication policy can explore more diverse communication options and adjust to a decentralized structure, thereby increasing its robustness.}
% This paper proposes a decentralization-oriented communication policy training method {\tool}, which is robust against various abnormal conditions, for any existing communication policy. Technically, we define an adversary to identify and mask the critical communication channels. The policy learning of the adversary is modeled as a MARL problem based on the CTDE paradigm, with the %fundamental 
% optimization objective to minimize the difference between the expected reward and actual reward, where the reward function has been reconstructed. 
As a training method to improve robustness of existing policies, {\tool} can be combined with any learnable communication policy. Experimental results show that compared to baselines, {\tool} can achieve decentralized communication policy and better improve the robustness and performance of communication policies.

\bibliographystyle{named}
\bibliography{ref}

\end{document}